\documentclass[longauth]{aa} 
\usepackage{pdflscape} 
\usepackage{lscape} 
\usepackage{graphicx} 
\usepackage{txfonts} 
\usepackage{booktabs} 
\usepackage{natbib} 
\bibpunct{(}{)}{;}{a}{}{,} 
\usepackage[colorlinks,linkcolor=blue,citecolor=blue,urlcolor=blue]{hyperref} 
\usepackage{cleveref} 

\def\NclusRASS{12,247} 
\def\NclusRASSused{4,303}
\def\NclusRASSusedvisual{3,946} 

\def\MannrelaxedclustersPct{28.1}
\def\relaxedclustersPctSeppi{29.1}
\def\relaxedclustersPctSeppiOPTcenter{30.8}
\def\BCGunderfiveKpcClusters{41}

\def\BCGrelaxedclustersPct{\textcolor{black}{75.4}} 
\def\BCGunrelaxedclustersPct{\textcolor{black}{13.6}} 
\def\BCGXd {D$_{\rm BCG-X}$}
\def\BCGopt {D$_{\rm OPT-X}$}
\def\Rfive   {$r_{500}$}

\def\rhobBCG{$ 0.743_{-0.008}^{+0.008}$}
\def\rhoMBCG{$ 0.054_{-0.032}^{+0.033}$}
\def\rhozBCG{$ -0.032_{-0.048}^{+0.047}$}
\def\sigmazerobBCG{$ 0.021_{-0.001}^{+0.001}$}
\def\sigmazeroMBCG{$ -0.019_{-0.002}^{+0.002}$}
\def\sigmazerozBCG{$ 0.038_{-0.006}^{+0.006}$}
\def\DsigmazerobBCG{$ 0.511_{-0.009}^{+0.01}$}
\def\DsigmazeroMBCG{$ -0.26_{-0.035}^{+0.033}$}
\def\DsigmazerozBCG{$ 0.044_{-0.061}^{+0.062}$}

\def\rhobOPT{$ 0.74_{-0.01}^{+0.009 }$}
\def\rhoMOPT{$ 0.049_{-0.038}^{+0.037 }$}
\def\rhozOPT{$ -0.087_{-0.054}^{+0.05 }$}
\def\sigmazerobOPT{$ 0.012_{-0.002}^{+0.002}$}
\def\sigmazeroMOPT{$ -0.01_{-0.003}^{+0.003}$}
\def\sigmazerozOPT{$ 0.014_{-0.008}^{+0.008}$}
\def\DsigmazerobOPT{$ 0.215_{-0.005}^{+0.005}$}
\def\DsigmazeroMOPT{$ -0.156_{-0.014}^{+0.014}$}
\def\DsigmazerozOPT{$ 0.057_{-0.023}^{+0.023}$}

\begin{document}

   \title{eROSITA clusters dynamical state and their impact on the  BCG luminosity}

   \author{A. Zenteno\inst{1}, M. Kluge,\inst{2} R. Kharkrang,\inst{3} 
D. Hernandez-Lang,\inst{4,5} G. Damke,\inst{1} A. Saro,\inst{3,6,7,8,9,10}  R. Monteiro-Oliveira,\inst{11}  E. R. Carrasco,\inst{12}  M. Salvato,\inst{2} J. Comparat,\inst{2} M. Fabricius,\inst{2} J. Snigula,\inst{2} P. Arevalo,\inst{13,14}  H. Cuevas,\inst{15} J.L. Nilo Castellon,\inst{15} A. Ramirez,\inst{15} S.~V\'eliz Astudillo,\inst{15} M. Landriau,\inst{16} A.~D.~Myers,\inst{17} E. Schlafly,\inst{18} F. Valdes,\inst{19} B. Weaver,\inst{19}  J. J. Mohr,\inst{4,5} S. Grandis,\inst{20}  M. Klein,\inst{5} A. Liu,\inst{2} E. Bulbul,\inst{2} X. Zhang,\inst{2} J.~S. Sanders,\inst{2} Y. E. Bahar,\inst{2} V. Ghirardini,\inst{2} M. Ramos,\inst{2} \and F. Balzer\inst{2}
    }

   \institute{Cerro Tololo Inter-American Observatory, NSF’s National Optical-Infrared Astronomy Research Laboratory, Casilla 603, La Serena, Chile \email{alfredo.zenteno@noirlab.edu}
            \and Max Planck Institute for Extraterrestrial Physics, Giessenbachstrasse 1, 85748 Garching, Germany
            \and Universit\`e Paris-Saclay, CNRS, Institut d\'Astrophysique Spatiale, 91405, Orsay, France
            \and Max Planck Institute for Extraterrestrial Physics, Giessenbachstrasse 1, 85748 Garching, Germany
            \and Faculty of Physics, Ludwig-Maximilians-Universit\"at München, Scheinerstr. 1, 81679 Munich, Germany
            \and INAF - Osservatorio Astronomico di Trieste, via G. B. Tiepolo 11, 34143 Trieste, Italy
            \and IFPU - Institute for Fundamental Physics of the Universe, Via Beirut 2, 34014 Trieste, Italy
            \and Astronomy Unit, Department of Physics, University of Trieste, via Tiepolo 11, 34131 Trieste, Italy
            \and INFN - National Institute for Nuclear Physics, Via Valerio 2, I-34127 Trieste, Italy
            \and ICSC - Italian Research Center on High Performance Computing, Big Data and Quantum Computing, Italy
            \and Institute of Astronomy and Astrophysics, Academia Sinica, Taipei 10617, Taiwan
            \and International Gemini Observatory/NSF NOIRLab, Casilla 603, La Serena, Chile
            \and Instituto de Física y Astronom\'ia, Universidad de Valpara\'iso, Gran Breta\~na 1111, Valpara\'iso, Chile
            \and Millennium Nucleus on Transversal Research and Technology to Explore Supermassive Black Holes (TITANS), Chile
            \and Departamento de Astronom\'ia, Universidad de La Serena, Av. Ra\'ul Bitr\'an 1305, La Serena, Chile
            \and Lawrence Berkeley National Laboratory, 1 Cyclotron Road, Berkeley, CA 94720, USA
            \and Department of Physics \& Astronomy, University  of Wyoming, 1000 E. University, Dept.~3905, Laramie, WY 82071, USA
            \and Space Telescope Science Institute, 3700 San Martin Drive, Baltimore, MD 21218, USA
            \and NSF's National Optical/Infrared Research Laboratory (NOIRLab), 950 N. Cherry Ave, Tucson, AZ 85732, USA
            \and Universit\"at Innsbruck, Institut f\"ur Astro- und Teilchenphysik, Technikerstr. 25/8, 6020 Innsbruck, Austria
    }

   \date{Received September 30, 2024; accepted March 26, 2025}

    \titlerunning{Dynamical state of eRASS1 clusters}
    \authorrunning{A. Zenteno, M. Kluge,  R. Kharkrang et al.}

  \abstract
   {\textcolor{black}{The SRG (Spectrum Roentgen Gamma)/}eROSITA first public release contains \NclusRASS\ clusters and groups from the first 6 months of operation. We use the offset between the Brightest Cluster Galaxy (BCG) and the X--ray peak (\BCGXd) to classify the cluster dynamical state of \NclusRASSusedvisual\  galaxy clusters and groups.}
   {We aim to investigate the evolution of the merger and relaxed cluster distributions with redshift and mass, and their impact on the BCG.}
   {We use the X--ray peak from the eROSITA survey and the BCG position from the DECaLS DR10 optical data, which includes the DECam eROSITA Survey optical data, to measure the \BCGXd\ offset. We model the distribution of \BCGXd, in units of \Rfive, as the sum of two Rayleigh distributions representing the cluster's relaxed and disturbed populations, and explore their evolution with redshift and mass.   To explore the impact of the cluster's dynamical state on the BCG luminosity, we separate the main sample according to the dynamical state. We define clusters as relaxed if \BCGXd < 0.25\Rfive, \textcolor{black}{disturbed} if \BCGXd>0.5\Rfive, and as `diverse' if 0.25< \BCGXd<0.5.}
   {We find no evolution of the merging fraction in redshift and mass. We observe that the width of the relaxed distribution to increase with redshift, while the width of the two Rayleigh distributions decreases with mass. 
   The analysis reveals that BCGs in relaxed clusters to be brighter than BCGs in both the disturbed and diverse cluster population. The most significant differences are found for high mass clusters at higher redshift.}
   {The results suggest that BCGs in low-mass clusters are less centrally bound than those in high-mass systems, irrespective of dynamical state. Over time, BCGs in relaxed clusters progressively align with the potential center. This alignment correlates with their luminosity growth relative to BCGs in dynamically disturbed clusters, underscoring the critical role of the clusters dynamical state in regulating BCG evolution.
    }


\maketitle
%

\section{Introduction}

The process governing the build-up of the universe's large-scale structure is hierarchical, with larger formations emerging from successive mergers of smaller entities \citep{nelson24}. Given the continuous nature of this process, galaxy clusters exist in various dynamic states, ranging from those already virialized to highly disturbed clusters still undergoing formation \citep[e.g.,][]{Pandge19,Lourenco20,Yoon20,Monteiro-Oliveira22b, Abriola24}. Each of the cluster's evolutionary stages provides us with a different aspect of cluster science. For example, (almost) relaxed clusters are the most suitable targets for investigations into intra-cluster gas cooling/heating processes \citep[e.g.,][]{Soja18,Ueda20,Ueda21,Hlavacek-Larrondo22,Ruszkowski23}. They are also optimal for establishing scaling relations between different observables \citep[e.g.,][]{Monteiro-Oliveira21,Lovisari22, Bahar22, Chiu23, Doubrawa23}, determining cluster counts for cosmology probes \citep{Costanzi21, Bocquet24, Fumagalli24,ghirardini24,artis24}, and testing the validity of the Cold Dark Matter ($\Lambda$CDM) model \citep{Chan20,Brouwer21,Tam23}.  \textcolor{black}{Research topics that can be explored using} disturbed clusters \textcolor{black}{include} the investigation of how each cluster's constituents (gas, galaxies, and dark matter) behave under extreme circumstances \citep[e.g.,][]{Molnar16,Monteiro-Oliveira22a, Doubrawa20, Moura21, Albuquerque24,veliz24}, determination of the conditions for particle acceleration \citep[e.g.,][]{Botteon19}, and tests of exotic forms of dark matter \citep[e.g.,][]{bulbul14,Harvey15,Fischer23,sirks24}.

While the impact of galaxy mass and/or environmental factors on galaxy properties is relatively well understood \citep[e.g.,][]{Dressler80,peng10}, the influence of processes at larger scales on the evolution of galaxies, such as the highly energetic cluster mergers \citep[\textcolor{black}{up to} $10^{64}$ erg;][]{Sarazin04}, remains unclear. Given that the most pronounced effects of cluster collisions (e.g., gas heating and shock propagation) occur within a few million years after the pericentric passage \citep[e.g.,][]{Markevitch07,ZuHone11}, during a period of significant dynamical disruption \citep[e.g.,][]{Machado15b, Ruggiero19}, the observation of young post-collision systems can offer ``real-time'' insights into the potential effects on the stellar activity of the galaxies.

Pilot studies on individual mergers show a variety of results on the impact of mergers on the galaxy population. For example, \cite{hernandez-lang22} analysed the merging cluster SPT-CL~J0307-6225 using MUSE data. At redshift z = 0.58, they found that \textcolor{black}{most} of the emission-line galaxies lie close to the X--ray peak position\textcolor{black}{;} however when separating by galaxy population, they found that a third of the emission-line galaxies correspond to red-sequence cluster galaxies. Moreover, these emission-line red-sequence galaxies were classified as short-starburst galaxies, which, considering the peculiar velocities, were accreted before the merger between the two clusters occurred. On the other hand, 75\% of the blue emission-line galaxies have peculiar velocities that suggest they are in the process of being accreted. 
The young \citep[$0.5$ Gyr;][]{Machado13} binary merging cluster A3376 \citep[$z=0.046$;][]{Monteiro-Oliveira17b} displays a pair of radio relics \citep{Akamatsu12} as a result of the shock propagating through the intracluster medium. \cite{Kelkar20} concluded that this shock had distinct effects on high- and low-mass spiral galaxies. While the high-mass spirals were not significantly affected, keeping a stellar activity similar to that of the relaxed clusters, the low-mass ones had their stellar formation suppressed due to the interaction between the clusters. A plausible scenario is that the impact of the merger is temporary and, therefore, more evident in extremely young systems. \cite{Wittman24}, in a recent analysis of H$\alpha$ emitters distribution in 12 merging clusters, noted a slight trend indicating that the density of star-forming galaxies in the central region is lower compared to other galaxies in systems observed within 0.2 Gyr after the pericentric passage. However, this inference relies on \textcolor{black}{only} three systems, lacking sufficient statistical significance for validation.

In the quest of finding compelling evidence, authors have started to build larger samples and explore different dynamical state proxies. 
In the X--ray regime \cite{lourenco23} analysed  52 X--ray selected clusters, at a redshift range of 0.04-0.07, with different dynamical state to explore the effect of extreme ram pressure on the merging cluster's galaxy population. In particular, they look at the prevalen\textcolor{black}{ce} of jellyfish galaxies as a function of the cluster dynamical state, unable to find a correlation between them. 
\cite{wenhan15} used an optical proxy to classify the cluster dynamical state of 2092 optically selected rich galaxy clusters, within a redshift range of $0.05 < z < 0.42$. The study explored the impact of the cluster dynamical state on the bright end of the luminosity function. They found the knee of the Schechter function ($m^*$) to be fainter in relaxed clusters with respect to disturbed clusters, while the BCG being brighter.  
\citet[][Z20 henceforth]{zenteno20} used SZ, X--ray, and optical information to classify the dynamical state of clusters and to study its impact on the cluster's luminosity function. Z20 used the offset between the BCG and the SZ centroid and X--ray peak to classify the dynamical \textcolor{black}{state} of 288 massive SPT-SZ selected clusters \citep{bleem15}. Z20 found that i) the faint end of the luminosity function is steeper, ii) $m^*$ is brighter, and iii) the BCG is fainter for merging clusters in comparison to relaxed clusters at $z \gtrsim \textcolor{black}{0.5}$, most significantly at $0.5 \times$ R$_{200}$ when using galaxies of all colors.  \cite{aldas23} expanded this work exploring the impact of the mergers on the red sequence, finding merging cluster\textcolor{black}{s} having a broader red sequence also at the same $z \gtrsim \textcolor{black}{0.5}$ redshift range, and no difference at lower redshifts. Both work shows that at the top of the cluster mass function, at $z \gtrsim \textcolor{black}{0.5}$, quenching happens in-situ, while at lower redshift quenching happens ex-situ.

As the impact of the cluster dynamical state on galaxy populations starts to become clearer, efforts to both improve dynamical state proxies and to enlarge cluster samples with dynamical state classification have been carried out. That includes theoretical \citep[e.g.,][]{cui17,capalbo21,deluca21} as well as observational work \cite[e.g.,][]{yuanhan20}. In particular, \cite{yuanhan20} used archival {\it Chandra} data of 964 clusters to explore new dynamical state proxies. They used the concentration index c, the centroid shift $w$, and the power ratio to create \textcolor{black}{their} own two proxies. They found that such new proxies, a profile parameter and an asymmetry factor, are excellent indicators to classify the cluster dynamical state. In a follow\textcolor{black}{-}up study \cite{yuanhan22} expanded the sample using 1,308 XMM and 22 new {\it Chandra} X--ray data, bringing the sample to \textcolor{black}{1,844} clusters, finding agreement between  XMM and {\it Chandra} data. Other studies on X-ray–selected samples have utilized data from the extended ROentgen Survey with an Imaging Telescope Array (eROSITA) aboard the Spectrum-Roentgen-Gamma (SRG) mission \citep{merloni12,predehl21,merloni24}. \cite{ghirardini22} analysed X-ray data from eROSITA to measure eight morphological parameters for 325 galaxy clusters and groups \citep{liu22}. This dataset, covering approximately 140 deg$^2$ from the eROSITA Final Equatorial-Depth Survey (eFEDS), showed that the fraction of relaxed objects is around 30–35\% and found no evidence of redshift evolution.

Another commonly used proxy for classifying the dynamical state of a large number of galaxy clusters is the BCG-X-ray peak/centroid offset (\BCGXd). This metric has been widely adopted as a reliable indicator of cluster dynamical states \citep[e.g.,][]{mann12, song12b, lopes18, zenteno20}. The \BCGXd\ reflects the expectation that the collisionless BCG aligns with the bottom of the dark matter (DM) potential, while the collisional gas is displaced relative to the DM and galaxy components. Moreover, since this method relies on shallow optical and X-ray imaging, it enables measurements to be conducted without requiring a substantial investment of telescope time, per cluster.

In this work we go further and provide a dynamical state classification for \NclusRASSusedvisual\  uniformly X--ray selected clusters. \textcolor{black}{We use a sample of clusters and groups from the first eROSITA All-Sky Survey \citep[eRASS1;][]{bulbul24,kluge24}}, and combine that information with optical data from the Legacy Survey DR10 \citep{dey19}\textcolor{black}{, which includes the DECam eROSITA Survey (DeROSITAS)}. We use the dynamical state information to understand its impact on the Brightest Cluster Galaxy.

This paper is organized as follows. In \S~\ref{sec:xray} we introduce the X--ray data and cluster sample.  In \S~\ref{sec:optical} we describe the optical data and the BCG selection we use paired with the \textcolor{black}{X}--ray data. In \S~\ref{sec:offsetdistribution} we explore \textcolor{black}{the BCG-X--ray offset distribution} (\BCGXd). In \S~\ref{sec:dynamicalstate} we use the \BCGXd\ to classify the cluster dynamical state, provide examples, and analyze the luminosity offset between the BCG luminosity and $m^*$ at the cluster redshift.  Finally, in \S~\ref{sec:conclusions} we present our conclusions.

Throughout this work, we assume a flat $\Lambda$CDM cosmology with $H_0$=68.3 km s$^{-1}$ Mpc$^{-1}$ and $\Omega_{\rm M}$ = 0.299 \citep{bocquet15}.

\section{Data}
\label{sec:data}
\subsection{eRASS1 X--ray data}
\label{sec:xray}
The first eROSITA All-Sky Survey \citep[eRASS1;][]{merloni24} was undertaken between December 2019 and June 2020, and covers the western Galactic hemisphere (Galactic longitude $359.9442\degr > l > 179.9442\degr$) with variable exposure time. In equatorial coordinates, this \textcolor{black}{primarily} corresponds to the southern sky.
With a moderate angular resolution of $\sim$30\arcsec, galaxy clusters are resolved as extended sources\textcolor{black}{, though  8,347 clusters of galaxies have been identified in the point-sources catalog  \citep{balzer25}}. The X--ray positional uncertainty is described by \texttt{RADEC$\_$ERR}. \texttt{RADEC$\_$ERR} corresponds to the sum of the errors in R.A. and Declination in quadrature, which for point sources 99\% are within 10\arcsec\ \citep{merloni24}. 

The eRASS1 identified galaxy cluster catalog \citep{bulbul24,kluge24}, contains \NclusRASS\ clusters and groups that were selected in the X-ray, and identified at optical and near-infrared wavelengths. The latter step made use of the Legacy Surveys' DR9 and DR10 source catalogs, which exclude the Galactic plane (Galactic latitude $|b|\lesssim20\degr$) and thereby reduce the usable eRASS1 survey area to 13,116\,deg$^2$.

Cluster candidates are optically identified when at least two galaxies are found at any given redshift and the richness of the cluster candidate is $\lambda > 3$. For this procedure, the tool eROMaPPer \citep{chitham20} is used, which is based on the redMaPPer algorithm \citep{rykoff12,rykoff14,rykoff16}.

The X-ray properties of the eRASS1 galaxy clusters are published in \cite{bulbul24} and the optical properties in \citet[][K24 henceforth]{kluge24}. The catalog has a known statistical contamination of $\bar{P}_{\rm cont}=14\%$. Most of these contaminants have low richness and \textcolor{black}{either} low or high redshifts. We limit our analysis to a higher-quality subsample by applying further constraints:
\begin{itemize}
    \item the redshift of the cluster is greater than $z=0.05$ (BEST\_Z $> 0.05$),
    \item the photometric redshift is smaller than the local limiting redshift (IN\_ZVLIM=True),
    \item the normalized richness is greater than 16 (LAMBDA\_NORM > 16), corresponding to clusters with mass $M_{\rm 200m}\gtrsim10^{14}$\,M$_\odot$,
    \item the probability of the cluster being a contaminant is less than 10\% (PCONT < 0.1),
    \item the fraction of the cluster area masked is under 10\% (MASKFRAC < 0.1), and
    \item the cluster mass is measured \textcolor{black}{$M_{500} > 10^{14}$M$_\odot$}.
    \item \textcolor{black}{$R_{500}$ > 0.}
\end{itemize}
These selection criteria reduce the number of clusters in our sample to \NclusRASSused. We will apply a further criterion on the quality of the BCG choice in Sec. \ref{sec:bcg_selection}. \textcolor{black}{The final sample has a purity of 98\% (1-PCONT = 0.98)}. As the X--ray center, we use the refined measurement RA\_XFIT and DEC\_XFIT in \citet{bulbul24}.

\subsection{Optical data\textcolor{black}{: DECaLS DR10 \& DeROSITAS} }
\label{sec:optical}

For the BCG position and brightness, K24 used the Legacy Survey DR9 and DR10 public release\footnote{https://www.legacysurvey.org/dr10/description/} \citep{dey19}.  DR10 was compiled using data from several programs, large and small, to fill the extragalactic sky, covering a large fraction of the western galactic hemisphere. Among those programs, most notably are the Legacy Survey DR9, the Dark Energy Survey \citep[DES;][]{abbott18}, The DECam Local Volume Exploration Survey \citep[DELVE;][]{drlica-wagner22}\footnote{https://delve-survey.github.io/}, and the DECam eROSITA Survey (DeROSITAS; PI: Zenteno)\footnote{https://noirlab.edu/science/programs/ctio/instruments/Dark-Energy-Camera/DeROSITAS}$^,$\footnote{http://astro.userena.cl/derositas/}.  In particular, our team designed DeROSITAS to fill the German eROSITA extragalactic sky to a $griz$ minimum depth of  22.7 (23.5), 23.2 (24.0), 23.3 (24.0), 22.5 (23.2) AB magnitudes at 10(5)$\sigma$, respectively. The goal was to reach a depth of $m^*+1$ at $z\sim 0.9$ to obtain precise cluster photometric redshifts \citep[see, for example,][]{song12b,klein18, klein19}, a key ingredient to realize eROSITA cluster cosmology \citep{ghirardini24}. DeROSITAS dedicated about 100 nights using the Dark Energy Camera \citep[DECam;][]{flaugher15} on the Cerro Tololo Inter-American Observatory Blanco 4m telescope, between the 2017A and 2022A semesters, to ``fill'' over 13,411 deg$^2$ of extragalactic eROSITA-DE sky. DeROSITAS defined extragalactic eROSITA-DE sky as the sky excluding areas at decl. $\gtrsim +30$, with high Galactic hydrogen column density (N$_{\mathrm{H}}$ $>10^{21}$ cm$^{-2}$), a high density of stars ($>$30,000 stars/deg$^2$), or strong Galactic dust reddening (E(B-V)$>$0.3 mag).  Observations were carried out using the Legacy Survey tiling scheme with the 4 dithering pattern used by \textcolor{black}{the} DELVE-WIDE survey \citep[][]{drlica-wagner21}. DeROSITAS used the effective exposure time scale factor $t_{\rm eff}$ \citep[][]{neilsen16} to repeat any observations with a $t_{\rm eff}$ less than 0.20 (meaning the observation was equivalent to a 20\% exposure time of the sky at the zenith on a dark night under ideal conditions). The use of $t_{\rm eff}$ also allowed DeROSITAS to adjust the required exposure time in a night by night basis; the total exposure time needed to reach full depth per pointing was estimated ($T_{\rm full depth} = \sum^{n=3}_{i=1} t_{\rm eff,i} \times T_{\rm exp,i}$) and distributed among the remaining dithered observing scripts, just before a DeROSITAS observing night. To reach a homogeneous depth, DeROSITAS observed no more than one dither per night. 
As at the time multiple surveys and programs were covering the southern sky, DeROSITAS worked in coordination with other PIs to avoid duplication of efforts (e.g., DELVE). A table with the observations can be found in Table~\ref{tab:observations}. 

\begin{table*}
    \small
	   \centering
        \caption{DeROSITAS observations.}
        \label{tab:observations}
	   \begin{tabular}{lcccccccc} 
         \hline 
         \hline
         program & N nights & $N_{\rm exp}$ & $u$ & $g$ & $r$ & $i$ & $z$ & $Y$    \\
                 & total    &               & hrs.  & hrs. & hrs.  & hrs.  & hrs.   & hrs.   \\
         \hline
    2017A-0388   &  8.5     & 1704          & 0 & 4.45 & 6.35 & 11.14 & 20.28 & 0.5  \\
    2018A-0386   &  7       & 4269          & 0 & 4.06 & 15.32 & 11.52 & 53.77 & 7.12  \\
    2019A-0272   &  13      & 1911          & 0 & 0 & 4.2 & 0.52 & 44.84 & 0 \\
    2019B-0323   &  18      & 1918          & 0 & 2.19 & 4.31 & 34.27 & 20.68 & 0  \\
    2020A-0399   &  19      & 1729          & 0 & 1.91 & 5.83 & 31.96 & 5.94 & 0   \\
    2020A-0909   &  4       & 775           & 11.66 & 0 & 0 & 3.48 & 0 & 0      \\
    2020B-0241   &  13.5    & 1657          & 1.13 & 1.98 & 1.6 & 20.89 & 17.31 & 1.53  \\
    2021A-0149   &  18.5    & 3584          & 0 & 4.58 & 2.26 & 76.46 & 7.19 & 36.82  \\
    2021A-0922   &  3       & 608           & 0 & 0.43 & 0 & 15.25 & 0.53 & 8.93   \\
    2022A-597406 &  1       & 270           &  0 & 0.26 & 0 & 5.59 & 0 & 6.29    \\
        \hline
    Total        &  107     & 18425         & 12.79 & 19.86 & 39.87 & 211.08 & 170.54 & 61.19 \\

        \hline
    \end{tabular}
    
\begin{minipage}{\linewidth}
	   \centering
\footnotesize
Includes only observations with $t_{eff}>0.2$. Under  conditions too poor for $griz$ $Y-$band observations were taken.
\end{minipage}

\end{table*}

DECaLS uses the community pipeline reduced data  \citep{Valdes14}, and {\it The Tractor} \citep{lang16}\footnote{https://github.com/dstndstn/tractor.} to perform the source photometry.  {\it The Tractor} used forward-modeling to perform source extraction on pixel-level data, modeling observed images as intrinsic profiles convolved with the point-spread function of each image.  DR10 intrinsic profiles include six morphological types: point sources (``PSF''), round exponential galaxies with a variable radius (``REX''), de Vaucouleurs (``DEV'') profiles for elliptical galaxies, exponential (``EXP'') profiles for spiral galaxies, Sersic (``SER'') profiles, and  ``DUP'' for Gaia sources coincident with extended sources (no flux is reported for this type). Those models are fitted to the data at the pixel level, simultaneously modeling all individual images overlapping a source. For photometry we use $mag_g$, $mag_r$, $mag_i$, and $mag_z$ while filtering out objects with ``type'' ``PSF'', ``DUP'', and negative flux in any band. Source morphological parameters are held fixed between the different filters, while source fluxes are allowed to vary between filters but are held constant in time. \textcolor{black}{DECaLS astrometry is fully aligned with Gaia Data Release 2 \citep[DR2;][]{gaia16,gaia18astro},  achieving astrometric residuals that are typically within $\pm$0.03\arcsec.}

\subsection{BCG Selection} 
\label{sec:bcg_selection}
Each of K24's BCG is selected as the brightest cluster member based on its $z$-band magnitude. We have verified that this selection is independent of the choice of filter. Shifting the selection wave band to the filter redwards the 4,000\AA\ break would result in consistent BCG choices in 98\% of the cases.

The search radius around the X--ray detection is the cluster radius $R_{\lambda}$ \citep[][K24]{rykoff14}
\begin{equation}
    R_{\lambda} = 1.0h^{-1} {\rm Mpc} (\lambda/100)^{0.2}.
\end{equation}

K24 expect that the BCG choice is consistent in only $80\sim85\%$ of the cases with other selection criteria based on the presence of diffuse intracluster light or visual inspection. To further clean our cluster sample we visually inspected all \NclusRASSused\ clusters and, assuming a conservative approach, we exclude \textcolor{black}{$\sim$8\%} of BCGs which are evidently wrong.  "Evidently wrong" BCGs includes, for example, cases where we find a larger and brighter galaxy, clearly visible close to the X--ray peak. Reasons for this selection error may include star-formation, dust absorption, or erroneous photometric measurements due to the  BCG's extended outer component. Those features could move the BCG off the red sequence. There are a couple cases where one entry had information from two overlapping clusters with the BCG of one cluster and the \Rfive\ from the other one, rendering a\textcolor{black}{n} erroneous large \BCGXd\ offset (for example, 1eRASSJ033101.2-522843). Those rare cases are described in the appendix A.3 in K24 where the best redshift type was changed to a literature redshift, becoming unreliable.  This reduces the BCG sample to \NclusRASSusedvisual\ which we use for our analysis. 
As an alternative proxy for the cluster center, we also use the optical center RA\_OPT and DEC\_OPT in K24 to compare to previous results found in \cite{seppi23}. RA\_OPT and DEC\_OPT correspond to the most probable cluster center found by a Bayesian classification algorithm \citep{rykoff14,seppi23}.

\subsection{BCG-X--ray offset distribution}
\label{sec:offsetdistribution}

Following previous works (\citealt{saro15}, Z20), we characterize the offset distribution as the sum of two contributions, one describing the cluster population with small \BCGXd\ and another one with large \BCGXd. 
In particular we fit a model that assumes that both the well-centered and disturbed population are well described by a Rayleigh distribution:
\begin{equation}
\label{eq:s15}
P(x|\rho, \sigma_0, \sigma_1) = x \left(\frac{\rho}{\sigma_0^2}e^{-\frac{x^2}{2\sigma_0^2}} + \frac{1-\rho}{\sigma_1^2}e^{-\frac{x^2}{2\sigma_1^2}}\right)
\end{equation}
where $x = r/$\Rfive, and $\rho$ ($1- \rho$) is the fraction of well-centred (disturbed) clusters with a variance of $\sigma^2_0$ ($\sigma^2_1$).

\begin{figure*}
	\includegraphics[width=\linewidth]{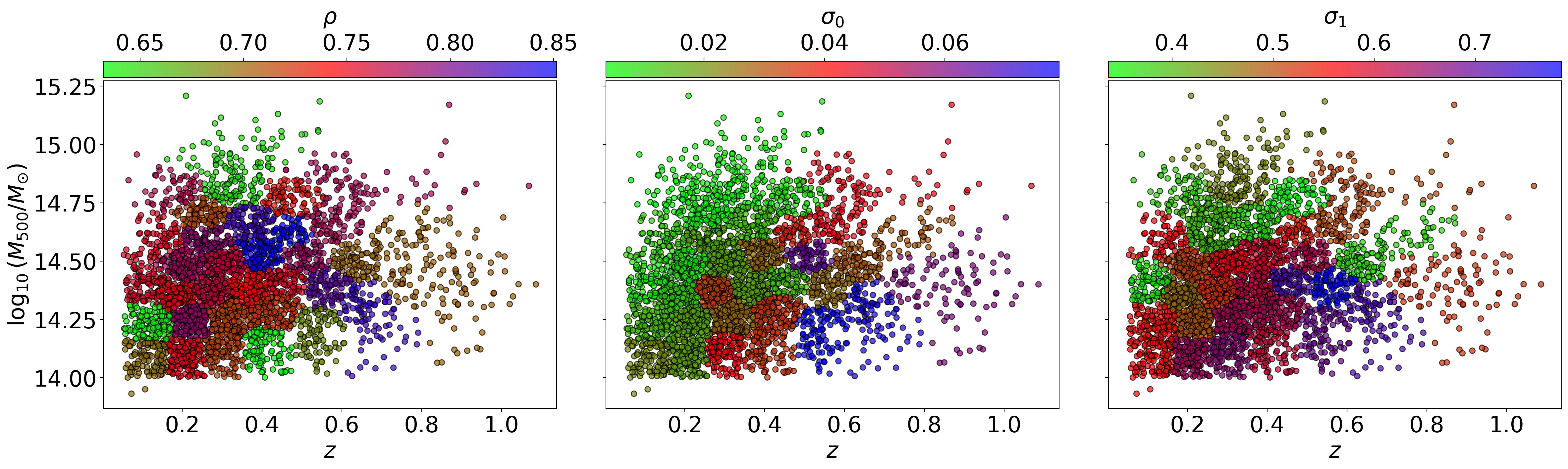}
    \caption{Best fit $\rho$, $\sigma_0$, and $\sigma_1$ parameters in bins of mass and redshift derived from Eq. \ref{eq:s15}. \textcolor{black}{For $\sigma$, trends are observed with both redshift and mass.}}
    \label{fig:scatterplot3par}
\end{figure*}

The large number of clusters allows us to explore the mass and redshift evolution. We thus divide the cluster sample in 35 mass and redshift bins (to obtain about 100 clusters per bin), with a K-means algorithm. In each individual bin, we thus fit the previously described miscentering model.  
As an example, we show in Fig. \ref{fig:scatterplot3par} the mass and redshift distribution of the analysed cluster catalog, where each of the analysed bins has been colour-coded according to the resulting best fit parameters derived within \textcolor{black}{our} model.

\begin{figure*}
	\includegraphics[width=0.99\linewidth]{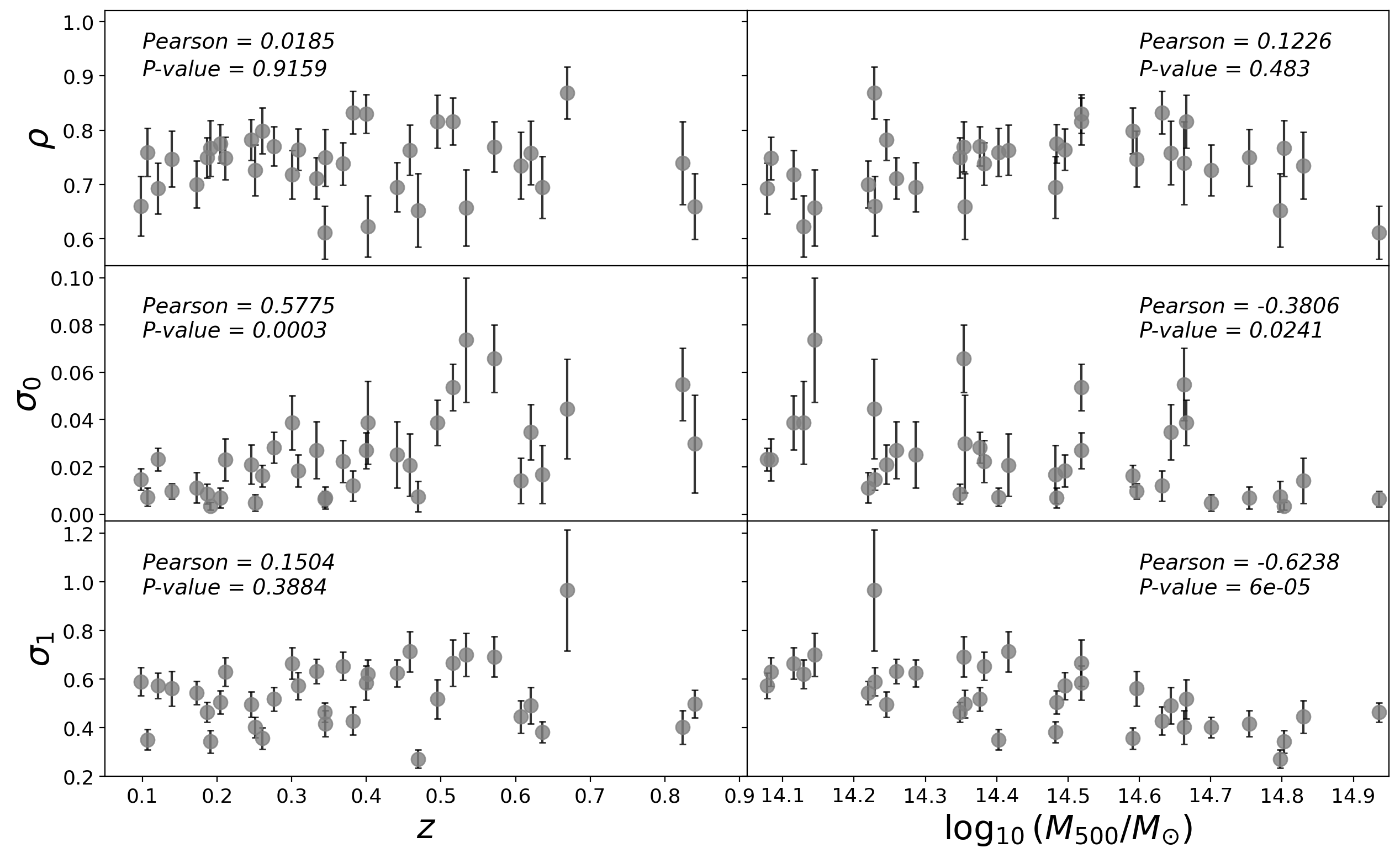}
    \caption{Marginalized distribution for $\rho$, $\sigma_0$, and $\sigma_1$ as a function of the \textcolor{black}{mean redshift and mean mass} of each analysed bin derived with Eq. \ref{eq:s15}. It can be see\textcolor{black}{n} that the fraction of (un)relaxed clusters ($1-\rho$) $\rho$ remains constant in redshift, while we find a positive correlation of $\sigma_0$ with redshift and a negative correlation of $\sigma_1$ with mass.}
    \label{fig:fit_all_bins}
\end{figure*}

Fig. \ref{fig:fit_all_bins} shows the marginalized posterior for each parameter as a function of the \textcolor{black}{redshift and mean mass} of the parent bin\footnote{We use the emcee algorithm \citep{emcee} to obtain the parameter space of the posterior distribution.}. The fraction of well centered clusters (described by the parameter $\rho$), does not appear to be significantly evolving with either mass or redshift. The trends highlighted by this figure are a correlation of $\sigma_0$ with redshift \textcolor{black}{and an anticorrelation with cluster mass,} and an anticorrelation of $\sigma_1$ with cluster mass.  In other words, well centred clusters have a tendency of being more disturbed at higher redshift, while smaller mass clusters \textcolor{black}{(groups)} have a tendency of having poorly centred BCGs (possibly due to their shallower potential well). 

\begin{figure}
	\includegraphics[width=\columnwidth]{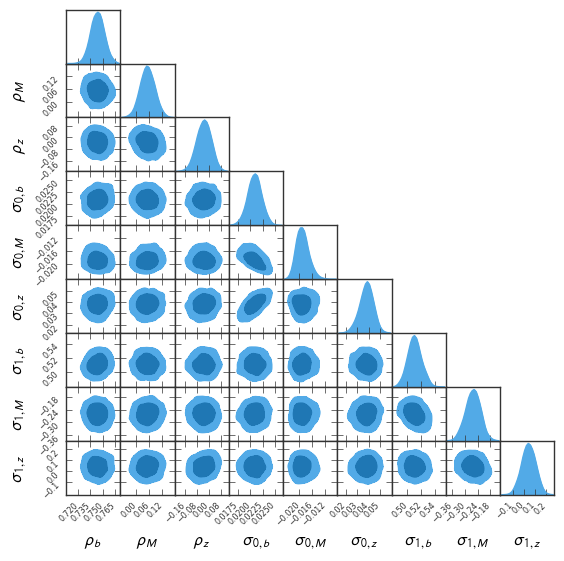}
    \caption{Marginalized distribution of the miscentering model described by Eq. \ref{eq:s15} and Eq. \ref{eq:gaussians_expansion}. $\sigma_0$ shows a positive redshift correlation ($\sigma_{0,z} =$ \sigmazerozBCG) \textcolor{black}{and a negative  correlation with mass ($\sigma_{0,M} =$ \sigmazeroMBCG).  $\sigma_1$ displays a negative  correlation with mass ($\sigma_{1,M} =$ \DsigmazeroMBCG)}. If we use the optical center instead of the BCG to measure the offset distribution the $\sigma_1$ positive redshift correlation ($\sigma_{0,z} =$ \sigmazerozOPT) and $\sigma_0$ negative mass correlation ($\sigma_{1,M} =$ \sigmazeroMOPT) stands\textcolor{black}{, albeit both much weaker}.}
    \label{fig:fullGTC}
\end{figure}

To delve on these results,  characterized by the apparent linear dependency of $\sigma_0$ and $\sigma_1$ with redshift and (log) mass highlighted by Fig. \ref{fig:fit_all_bins}, we expand our model to allow for mass and redshift dependence\textcolor{black}{, including the the X--ray peak positional error $P_{err}$  \citep[\texttt{RADEC$\_$ERR}; ][]{merloni24},} as follow:
\begin{equation}
\begin{split}
\rho &= \rho_b + \rho_z \times(z-\Bar{z}) + \rho_M \times {\rm log}_{10} (M_{500}/\Bar{M}_{500}), \\
s_0 &= \sigma_{0,b} + \sigma_{0,z}\times(z-\Bar{z}) + \sigma_{0,M}\times {\rm log}_{10} (M_{500}/\Bar{M}_{500}), \\
s_1 &= \sigma_{1,b} + \sigma_{1,z}\times(z-\Bar{z}) + \sigma_{1,M}\times {\rm log}_{10} (M_{500}/\Bar{M}_{500}), \\
\sigma_0 &= \sqrt{s_0^2+P_{err}^2},\\
\sigma_1 &= \sqrt{(s_0+s_1)^2+P_{err}^2}, 
\label{eq:gaussians_expansion}
\end{split}
\end{equation}
where $\Bar{z} = 0.296$ and $\Bar{M} = 10^{14.33} \textrm{M}_\odot$ are the median redshift and  $\textrm{M}_{500}$, use flat priors for $0\leq \rho\leq 1$, $\sigma_0 > 0$, and $s_1>0$
and $P_{err}$ in \Rfive\ units. $P_{err}$ evolves with redshift as it becomes a larger fraction of \Rfive\ at higher redshift. 
Fig. \ref{fig:fullGTC} shows the resulting posterior distribution for our expanded model (\citealt{bocquet_GTC}), confirming what is seen in Fig.~\ref{fig:fit_all_bins} (a positive redshift slope for $\sigma_{0,z}$ \textcolor{black}{and a negative mass slope for $\sigma_{0,M}$ and $\Delta 
 \sigma_M$}).  
\textcolor{black}{This confirms that well centered clusters seem to have a broader \BCGXd\ distribution at higher redshift, while lower mass clusters tend to have a broader \BCGXd\ distribution than their higher mass counterparts}.

\begin{figure}
	\includegraphics[width=0.96\linewidth]{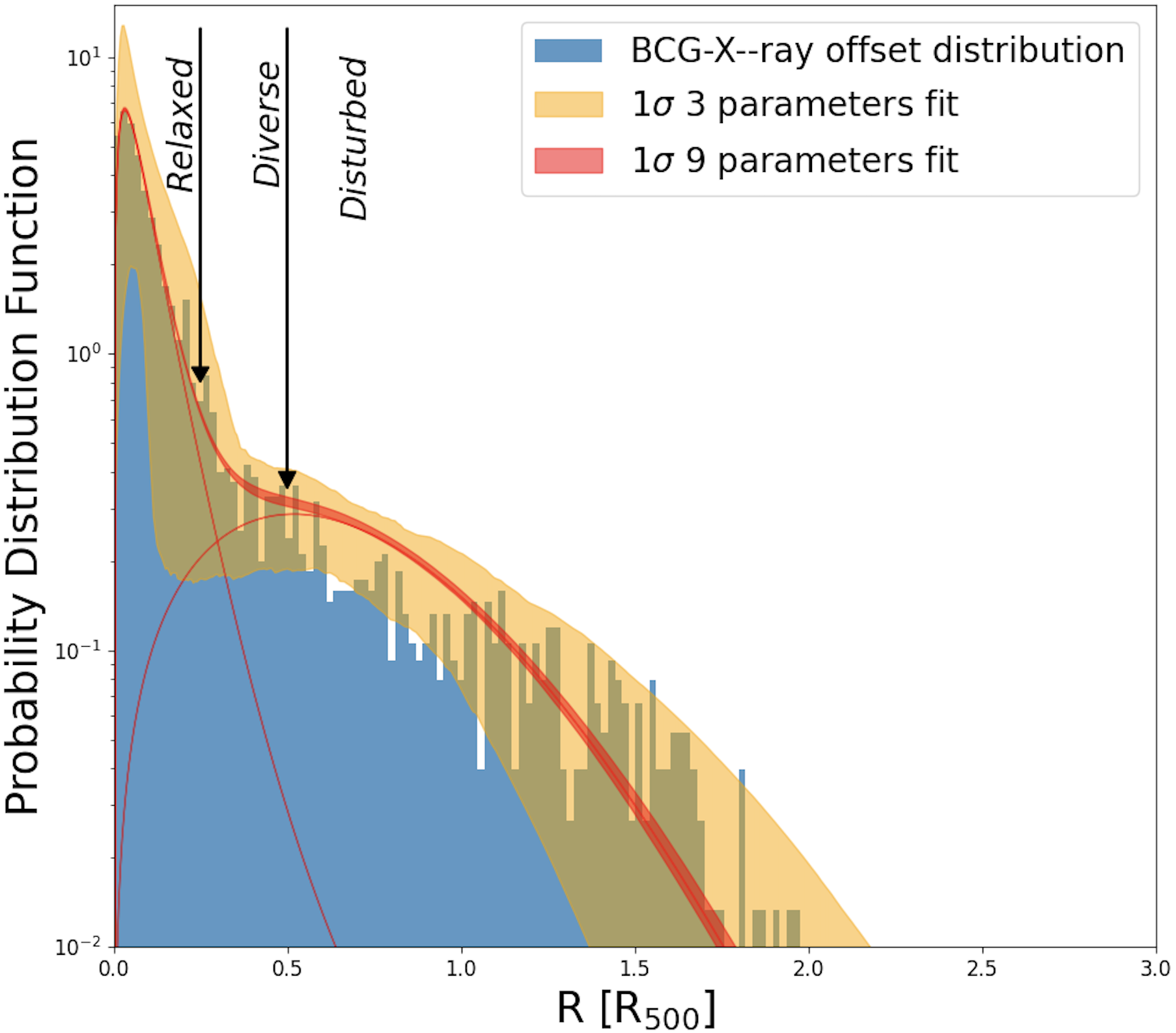}
    \caption{Observed X-ray/optical miscentering distribution (blue) for the analysed \textcolor{black}{total} cluster sample. The plot show\textcolor{black}{s} the results obtained using Eq. \ref{eq:s15}. Lighter \textcolor{black}{yellow} region show\textcolor{black}{s} the marginalized distribution for the \textcolor{black}{35} analysed bins \textcolor{black}{without evolutionary terms, while} solid red distributions show the results obtained using them (Eq. \ref{eq:gaussians_expansion}). \textcolor{black}{Also in red, the best fits for the relaxed and disturbed distributions components.}
    }
    \label{fig:distribution}
\end{figure}

Finally, we show in Fig. \ref{fig:distribution} the histogram of the observed miscentering distribution. The distribution expected from the fit of Eq. \ref{eq:s15} in each bin, marginalized over the entire sample, is shown in yellow, the results obtained using the 9 parameters expansion (Table~\ref{tab:params}) from equation~\ref{eq:gaussians_expansion} \textcolor{black}{are} shown in red. 

\begin{table}
	\centering
 \renewcommand{\arraystretch}{1.4} 
    \caption{9 parameters model \textcolor{black}{for the \BCGXd\ and  \BCGopt\ offsets. The most significant evolutionary terms are $\sigma_{0,M}$, $\sigma_{0,z}$, and $\sigma_{1,M}$ for \BCGXd\ and $\sigma_{0,M}$ and $\sigma_{1,M}$ for  \BCGopt.} }
    \label{tab:params}
	\begin{tabular}{lcc} 
         \hline 
         \hline
          Name variable & BCG center & optical center \\
         \hline
$\rho_b$ 	    & \rhobBCG        & \rhobOPT       \\
$\rho_M$ 	    & \rhoMBCG        & \rhoMOPT       \\
$\rho_z$ 	    & \rhozBCG        & \rhozOPT       \\
$\sigma_{0,b}$ 	    & \sigmazerobBCG  & \sigmazerobOPT \\
$\sigma_{0,M}$ 	    & \sigmazeroMBCG  & \sigmazeroMOPT \\
$\sigma_{0,z}$ 	    & \sigmazerozBCG  & \sigmazerozOPT \\
$\sigma_{1,b}$ & \DsigmazerobBCG & \DsigmazerobOPT\\
$\sigma_{1,M}$ & \DsigmazeroMBCG & \DsigmazeroMOPT\\
$\sigma_{1,z}$ & \DsigmazerozBCG & \DsigmazerozOPT\\
        \hline
    \end{tabular}
\end{table}

\subsection{eROSITA cluster dynamical state and BCG luminosity}
\label{sec:dynamicalstate}
\label{sub:bcglum}

To compare the BCGs luminosity as a function of the cluster's dynamical state we separate the \BCGXd\ offsets distribution in three samples. We redefine relaxed clusters as clusters with \BCGXd\ $< 0.25$ \Rfive, disturbed clusters as clusters with \BCGXd\ $> 0.5$ \Rfive, and leave everything in the middle as a diverse cluster population\textcolor{black}{, as shown in Fig.~\ref{fig:distribution}}. With this classification we find \BCGrelaxedclustersPct\% of the sample is classified as relaxed, while \BCGunrelaxedclustersPct\% as disturbed. 
If, on the other side, we follow \cite{mann12} classification then \MannrelaxedclustersPct\% are defined as relaxed, with a \BCGXd\ offset less than \textcolor{black}{42} kpc, with \BCGunderfiveKpcClusters\ clusters with \BCGXd\ offset less than 5kpc. This \textcolor{black}{result} is very similar when \cite{seppi23} criteria (\textcolor{black}{\BCGXd =} $\Delta_{\rm X-O} <$ 0.05 $\times$ R500$_c$) is used, finding a  \relaxedclustersPctSeppi\% of clusters relaxed. If we use the optical center (\BCGopt) instead of the BCG position, as used by \cite{seppi23}, then the number of relaxed clusters increases to \relaxedclustersPctSeppiOPTcenter\%, similar to the 31\% \cite{seppi23} found for the eFEDS sample. \textcolor{black}{\cite{ghirardini22} also characterized the dynamical state of clusters in the eFEDS sample. They separated the sample in\textcolor{black}{to} relaxed and disturbed clusters, using 7 X--ray morphological parameters, finding 30 to 35\% of them relaxed. This is consistent with \cite{seppi23} and our findings.}

\begin{figure*}
  \centering
  
  \begin{minipage}[t]{0.32\linewidth}
    \centering
    \includegraphics[width=\textwidth]{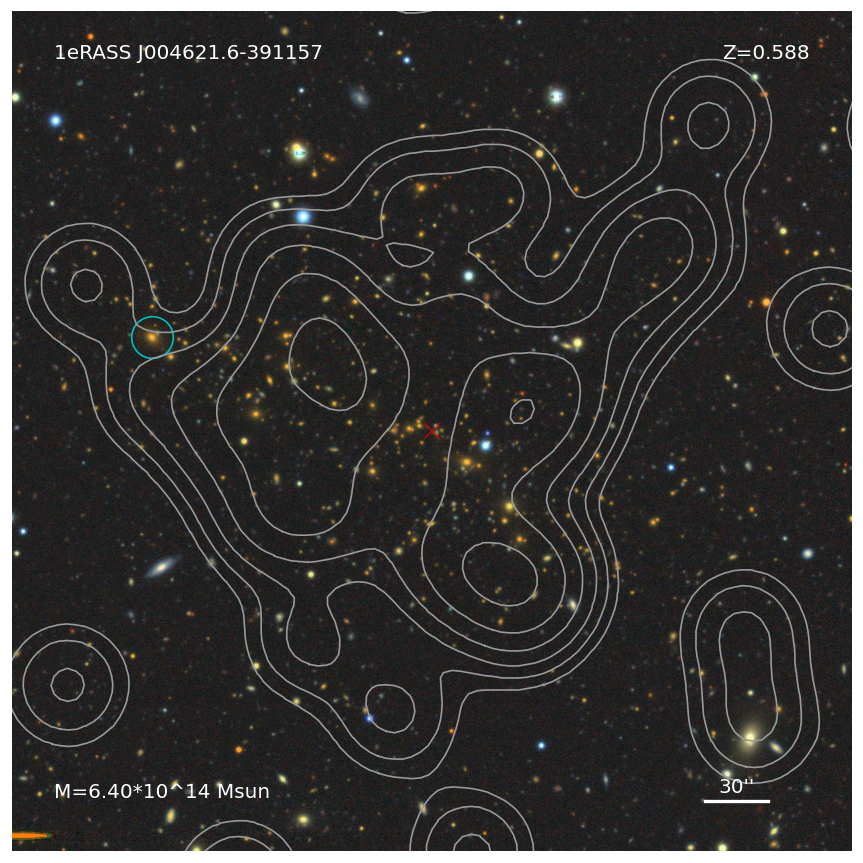} \\
  \end{minipage}\hfill
  \begin{minipage}[t]{0.32\linewidth}
    \centering
    \includegraphics[width=\textwidth]{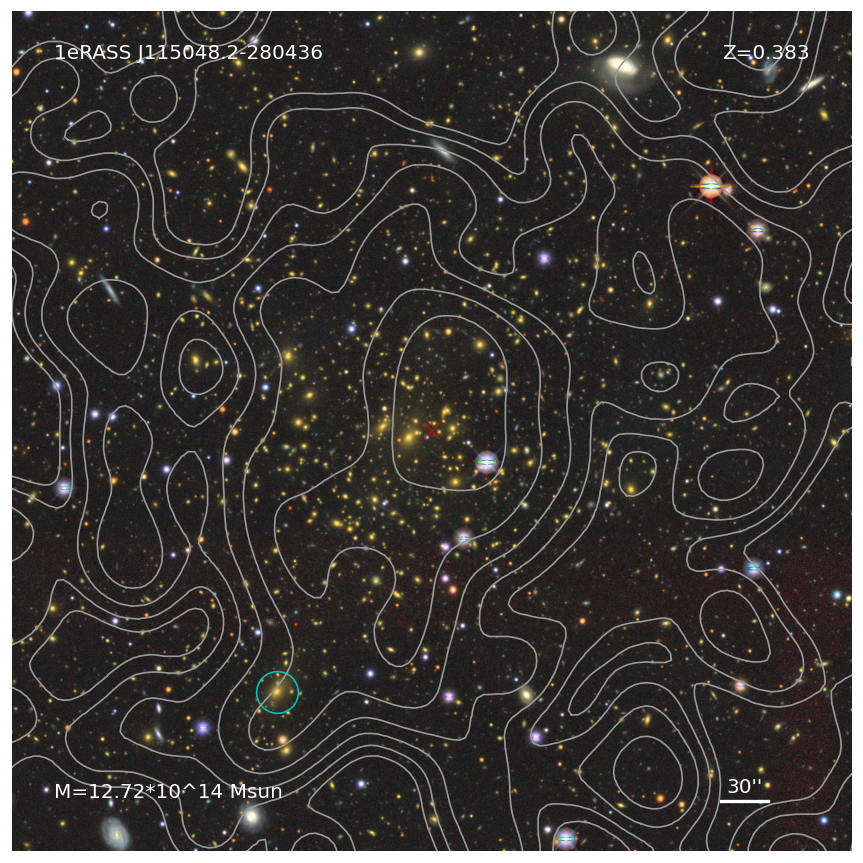} \\
  \end{minipage}\hfill
  \begin{minipage}[t]{0.32\linewidth}
    \centering
    \includegraphics[width=\textwidth]{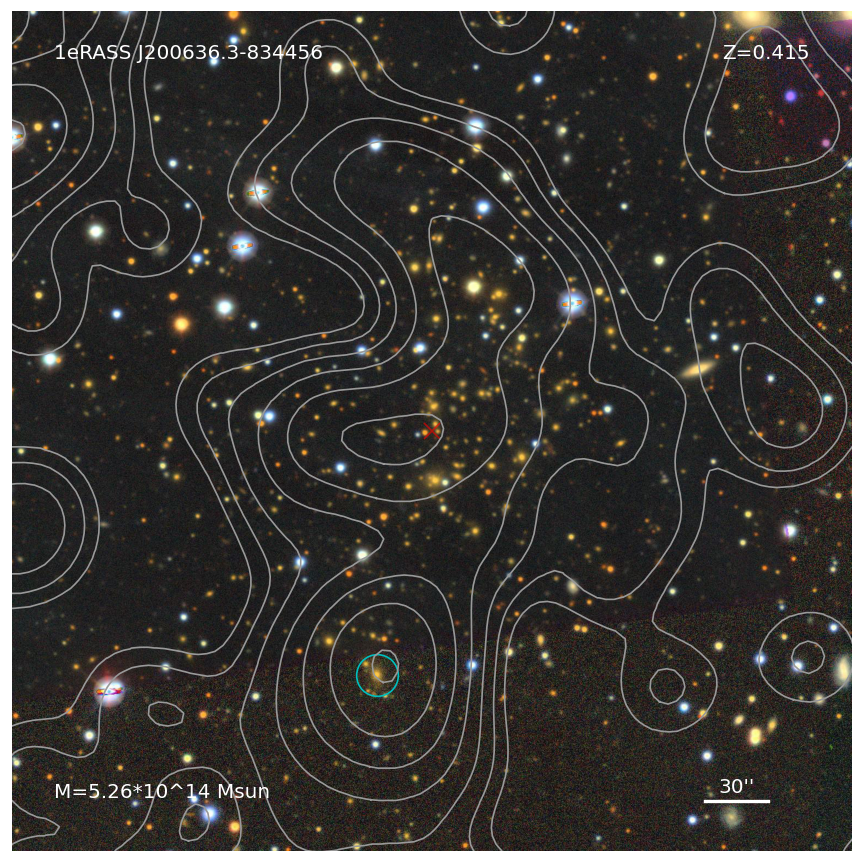} \\
  \end{minipage}
  
  \vspace{0.5cm} 
  
  \begin{minipage}[t]{0.32\linewidth}
    \centering
    \includegraphics[width=\textwidth]{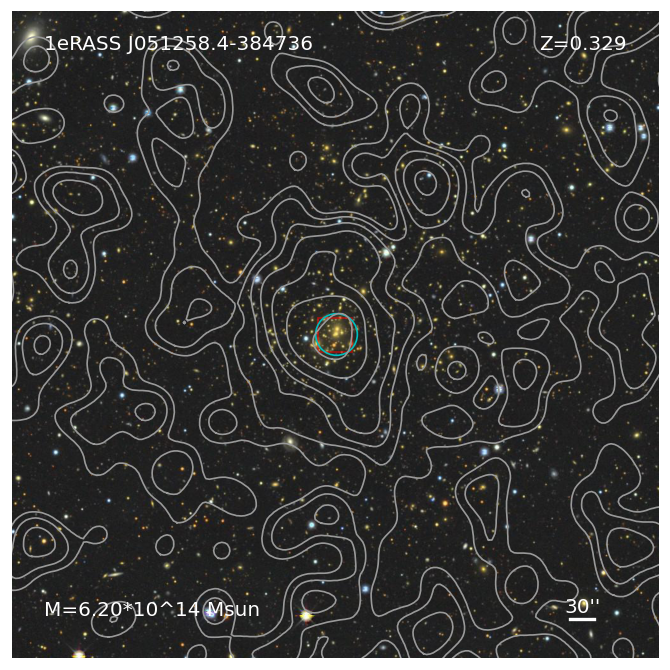} \\
  \end{minipage}\hfill
  \begin{minipage}[t]{0.32\linewidth}
    \centering
    \includegraphics[width=\textwidth]{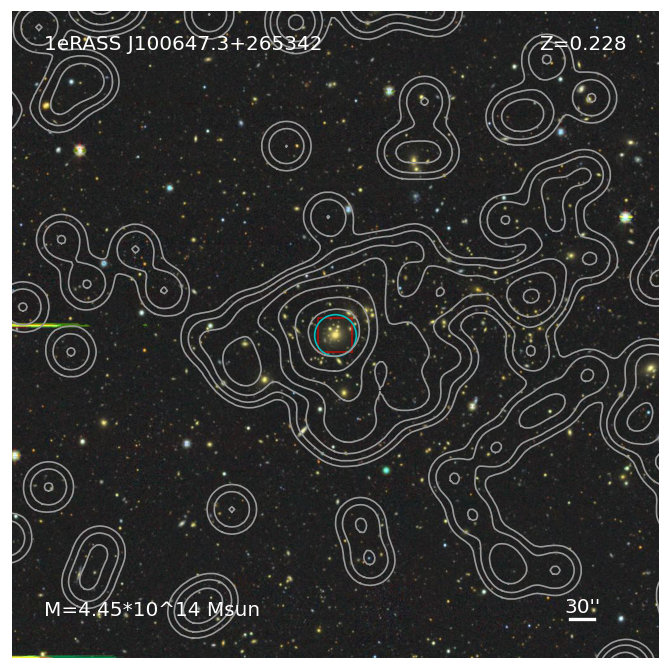} \\
  \end{minipage}\hfill
  \begin{minipage}[t]{0.32\linewidth}
    \centering
    \includegraphics[width=\textwidth]{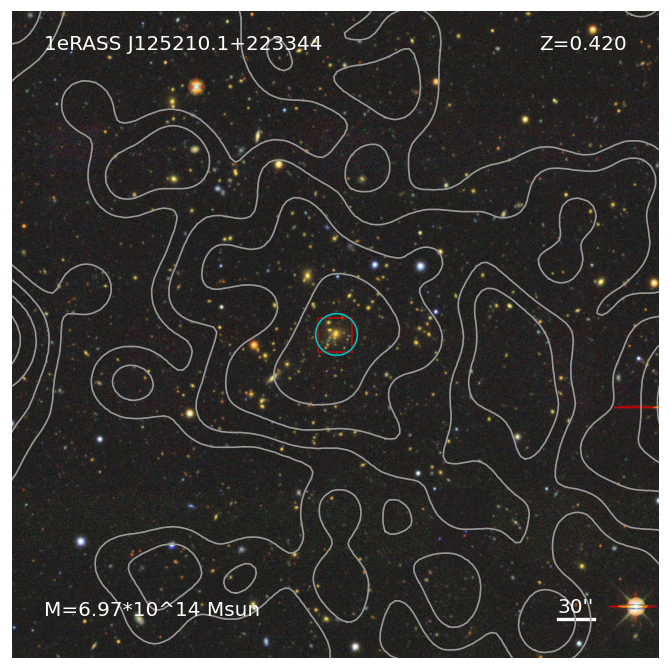} \\
  \end{minipage}
  \caption{{\it Top:} Example of three unrelaxed clusters as seen by DECaLS composite RGB images. {\it Bottom:} Example of three relaxed clusters as seen by DECaLS composite RGB images. Each image is centered at the peak of x-ray emission, shown by the red cross. white contours correspond to eROSITA X--ray contours. The cyan marker encircles the respective BCG of the cluster. The cluster redshift and cluster mass are indicated in the upper-right and lower-left corners of each panel, respectively, while the white bar shows the angular scale on the sky. North is up and East is left in all the images. Each image is 2.25 \Rfive\ side.}
\label{fig:relaxedunrelaxedclusters}
\end{figure*}

Examples of relaxed clusters are shown in \textcolor{black}{the bottom panels of} fig.~\ref{fig:relaxedunrelaxedclusters} while examples of disturbed clusters are shown in \textcolor{black}{the top panels of} fig.~\ref{fig:relaxedunrelaxedclusters}. We show tables for the \textcolor{black}{15} \textcolor{black}{of some of the} most relaxed and disturbed clusters in table~\ref{tab:relaxed} and table~\ref{tab:disturbed}, respectively.

\begin{table*}
\small
\caption{15 of the most relaxed eRASS1 clusters \& groups.}
\label{tab:relaxed}
\begin{tabular}{lccccccccc cc}
	    \hline
       \hline
        name & RASS1 R.A. & RASS1 Decl. & BCG R.A. & BCG  Decl. & $z$ & M$_{500}^a$ & R$_{500}^b$  & M$_{200}^c$ & R$_{200}^d$ & offset & offset\\
	    \hline
         eRASS1 & (J2000.0) & (J2000.0) & (J2000.0) & (J2000.0) & & $10^{14}  M_\odot$ & [$^\prime$]  & $10^{14}  M_\odot$ & [$^\prime$] & kpc & [$^{\prime\prime}$]\\
	    \hline
J005132.6-301946 &   12.8858 & $-30.3304$ & 12.8857 & $-30.33039$ & 0.263 & 4.84 & 4.45 & 7.12 & 6.87 & 1.28 & 0.31 \\
J022056.7-382850 &   35.2354 & $-38.4800$ & 35.2358 & $-38.47993$ & 0.228 & 6.60 & 5.57 & 9.76 & 8.61 & 4.28 & 1.16 \\
J031941.0-334444 &   49.9213 & $-33.7457$ & 49.9212 & $-33.74578$ & 0.411 & 5.61 & 3.28 & 8.37 & 5.09 & 2.28 & 0.41 \\
J041158.7-643624 &   62.9965 & $-64.6063$ & 62.9975 & $-64.60655$ & 0.155 & 4.25 & 6.68 & 6.19 & 10.28 & 4.88 & 1.79 \\
J044956.2-444019 &   72.4856 & $-44.6737$ & 72.4853 & $-44.67334$ & 0.152 & 5.28 & 7.33 & 7.72 & 11.29 & 4.02 & 1.51 \\
J051258.4-384736 &   78.2429 & $-38.7922$ & 78.2426 & $-38.79207$ & 0.329 & 6.20 & 4.04 & 9.23 & 6.25 & 4.62 & 0.96 \\
J100647.3+265342 &  151.6965 & $ +26.8952$ & 151.6965 & $ +26.89502$ & 0.228 & 4.45 & 4.86 & 6.52 & 7.50 & 2.42 & 0.65 \\
J102557.9+124108 &  156.4919 & $ +12.6857$ & 156.4916 & $ +12.68567$ & 0.143 & 7.74 & 8.76 & 11.40 & 13.53 & 2.76 & 1.09 \\
J114953.0+104639 &  177.4697 & $ +10.7772$ & 177.4695 & $ +10.77710$ & 0.308 & 5.06 & 3.98 & 7.49 & 6.15 & 3.69 & 0.80 \\
J115334.4+145117 &  178.3928 & $ +14.8544$ & 178.3930 & $ +14.85410$ & 0.230 & 6.91 & 5.61 & 10.23 & 8.68 & 4.79 & 1.29 \\
J125210.1+223344 &  193.0446 & $ +22.5611$ & 193.0445 & $ +22.56122$ & 0.420 & 6.97 & 3.47 & 10.46 & 5.40 & 3.10 & 0.55 \\
J135218.1+060536 &  208.0762 & $  +6.0934$ & 208.0761 & $  +6.09334$ & 0.491 & 4.40 & 2.65 & 6.57 & 4.12 & 2.65 & 0.43 \\
J152010.4-043844 &  230.0436 & $ -4.6456$ & 230.0438 & $ -4.64572$ & 0.434 & 6.21 & 3.26 & 9.32 & 5.06 & 4.66 & 0.82 \\
J202047.3-464621 &  305.1974 & $-46.7745$ & 305.1972 & $-46.77488$ & 0.196 & 3.81 & 5.25 & 5.56 & 8.09 & 4.79 & 1.46 \\
J224153.2-423549 &  340.4732 & $-42.5967$ & 340.4736 & $-42.59650$ & 0.206 & 4.18 & 5.21 & 6.12 & 8.03 & 4.40 & 1.29 \\
		\hline
\end{tabular}
\begin{minipage}{\linewidth}
\footnotesize
$^a$From K24. 
$^b$Calculated using M$_{500}$ = $\frac{4\pi}{3}$R$_{500}^3\rho_c(z)$. 
$^c$Estimated using \citet{duffy08} relation. 
$^d$Estimated similarly to R$_{500}$.
\end{minipage}
\end{table*}
\begin{table*}
	\centering
	\caption{15 of the most disturbed eRASS1 clusters \& groups}
	\label{tab:disturbed}
	\resizebox{\textwidth}{!}{\begin{tabular}{lccccccccc cc} 
	    \hline
       \hline
        name & RASS1 R.A. & RASS1 Decl. & BCG R.A. & BCG  Decl. & $z$ & M$_{500}$ & R$_{500}$ & offset & M$_{200}$ & R$_{200}$ & offset\\
	    \hline
         eRASS1 & (J2000.0) & (J2000.0) & (J2000.0) & (J2000.0) & & $10^{14}  M_\odot$ & [$^\prime$] & R$_{500}$ & $10^{14}  M_\odot$ & [$^\prime$] & R$_{200}$\\
	    \hline
J020628.4-145358 &   31.6187 & $-14.8998$ & 31.7063 & $-14.86412$ & 0.298 & 8.97 & 4.93 & 1.12 & 13.42 & 7.66 & 0.71 \\
J023301.0-713600 &   38.2554 & $-71.6006$ & 38.1995 & $-71.62224$ & 0.655 & 7.34 & 2.56 & 0.65 & 11.22 & 4.01 & 0.41 \\
J030345.7-775243 &   45.9482 & $-77.8773$ & 46.3799 & $-77.89506$ & 0.281 & 8.94 & 5.17 & 1.07 & 13.36 & 8.02 & 0.68 \\
J041145.4-001113 &   62.9355 & $ -0.1869$ & 62.8736 & $ -0.21061$ & 0.416 & 8.23 & 3.70 & 1.08 & 12.41 & 5.76 & 0.68 \\
J060008.8-200750 &   90.0346 & $-20.1373$ & 90.0701 & $-20.11553$ & 0.426 & 10.58 & 3.95 & 0.61 & 16.04 & 6.15 & 0.38 \\
J082944.9+382816 &  127.4372 & $ +38.4683$ & 127.4055 & $ +38.43778$ & 0.390 & 8.38 & 3.91 & 0.60 & 12.60 & 6.08 & 0.38 \\
J093512.7+004735 &  143.8022 & $  +0.7937$ & 143.8012 & $  +0.82560$ & 0.357 & 7.64 & 4.06 & 0.47 & 11.44 & 6.30 & 0.30 \\
J093521.3+023222 &  143.8402 & $  +2.5444$ & 143.8564 & $  +2.57652$ & 0.499 & 11.04 & 3.56 & 0.61 & 16.84 & 5.56 & 0.38 \\
J113808.7+275426 &  174.5363 & $ +27.9064$ & 174.5176 & $ +27.97732$ & 0.449 & 11.41 & 3.90 & 1.12 & 17.35 & 6.08 & 0.71 \\
J114140.7-190516 &  175.4197 & $-19.0874$ & 175.4817 & $-19.06264$ & 0.306 & 8.07 & 4.66 & 0.82 & 12.05 & 7.24 & 0.52 \\
J115048.2-280436 &  177.7044 & $-28.0810$ & 177.7356 & $-28.12728$ & 0.383 & 12.72 & 4.55 & 0.71 & 19.30 & 7.09 & 0.45 \\
J120106.8-395218 &  180.2710 & $-39.8722$ & 180.3294 & $-39.77250$ & 0.309 & 7.90 & 4.59 & 1.43 & 11.80 & 7.13 & 0.91 \\
J120316.9-213212 &  180.8234 & $-21.5347$ & 180.7842 & $-21.43674$ & 0.193 & 8.72 & 7.02 & 0.89 & 12.93 & 10.86 & 0.57 \\
J122838.4-363731 &  187.1588 & $-36.6248$ & 187.2220 & $-36.66564$ & 0.534 & 8.37 & 3.09 & 1.26 & 12.72 & 4.82 & 0.80 \\
J193739.9-600547 &  294.4181 & $-60.0974$ & 294.5110 & $-60.14393$ & 0.415 & 8.04 & 3.67 & 1.07 & 12.11 & 5.71 & 0.68 \\
		\hline
	\end{tabular}}
\end{table*}

For the BCG luminosity, we obtain the $griz$ BCG magnitudes (see \S~\ref{sec:optical}) by querying BCG positions\textcolor{black}{, obtained from K24,} from the NOIRLab DR10 DB. We match  95\% of the sample within 1$^{\prime\prime}$ radius, while for the remaining 5\% we use 0.5$^{\prime\prime}$ in the case of multiple entries and 2$^{\prime\prime}$ for the cases in which we find no match. For 197 BCGs one of the bands which straddle the 4,000 \AA\ break is missing, reducing the number of clusters with colors to 3,749.

To compare the BCGs brightness at different redshifts we model the $m^*$ evolution and subtract it from the cluster BCG luminosity (we use the magnitude corresponding to the band redward of the 4,000\AA\ break\footnote{we use $r-$band at $z<0.35$, $i-$band at $0.35<z<0.7$, and $z-$band at $z>0.7$}).  We model the $m^*$ evolution using a passively evolving composite stellar population \citep[see, e.g., ][]{song12b,zenteno16,hennig17}, with the help of  EZGAL \citep{mancone10}. We use the \cite{bruzual03} synthesis models, with \textcolor{black}{a} Chabrier initial mass function \citep{chabrier03}, assuming a formation redshift at $z = 3$ with an exponentially decaying star formation history with $\tau=0.4$ Gyrs. We use six distinct metallicities to match the tilt of the color-magnitude relation at low redshift. The metallicities correspond to  the $3L^*$, $2L^*$, $L^*$, $0.5L^*$, $0.4L^*$, and $0.3L^*$ luminosities \citep[we use a metallicity-luminosity relation parameters from ][]{poggianti01}. Finally we use $m^*$ from the Coma cluster \citep{iglesias-paramo03} to set the absolute normalization of the $m^*$ evolution.%

\begin{figure}
	\includegraphics[scale=0.57]{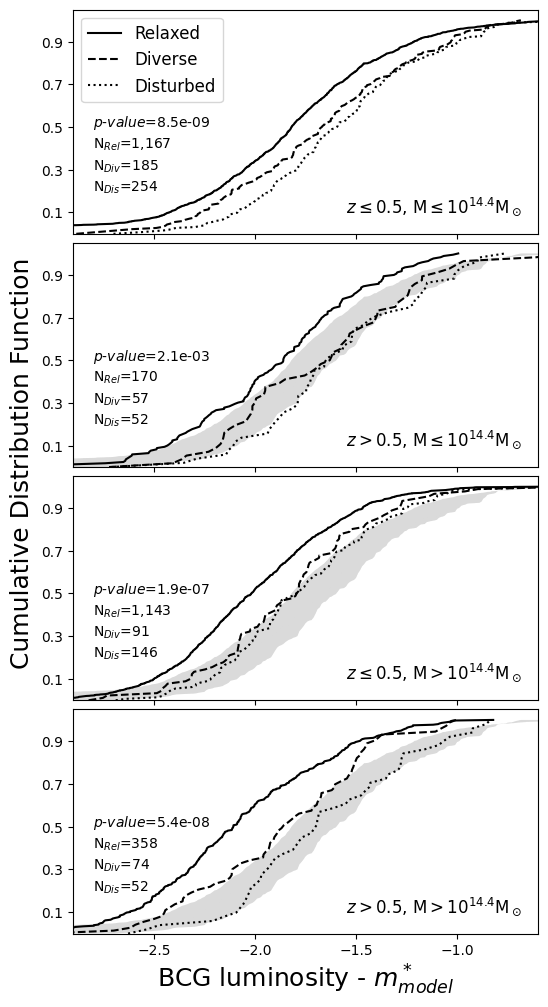}
    \caption{Example of cumulative distribution of the BCG luminosity. We use the BCG $r-$band for the $z<0.35$, $i-$band for $z<0.7$, and $z-$band for $z>0.7$ redshift range. The lines corresponds to relaxed (solid line; \BCGXd<0.25\Rfive),  disturbed (dotted line; \BCGXd>0.5\Rfive), and the rest (dashed line; 0.25\Rfive< \BCGXd<0.5\Rfive)  cluster samples. \textcolor{black}{In the bottom-left corner of each panel, the number of clusters used and the KS {\it p-value} comparing the relaxed and disturbed samples are displayed.} {\it Top panel}: BCGs at $z \leq \textcolor{black}{0.5}$ and cluster mass $\leq 10^{14.4}$M$_\odot$. {\it Middle-top panel}: BCGs at $z > \textcolor{black}{0.5}$ and cluster mass $\leq 10^{14.4}$M$_\odot$. {\it Middle-bottom panel}: BCGs at $z \leq \textcolor{black}{0.5}$ and cluster mass $> 10^{14.4}$M$_\odot$. {\it Bottom panel}: BCGs at $z > \textcolor{black}{0.5}$ and cluster mass $> 10^{14.4}$M$_\odot$. Shaded area corresponds to the area covered by the BCG distributions at the top panel. BCGs from the relaxed cluster sample tend to be brighter than BCGs in both, the disturbed and the general population samples, most significantly at higher redshifts.}
    \label{fig:BCGluminosity}
\end{figure}

We find all BCGs fainter than  $m^*-3.2$ while the average luminosity is \textcolor{black}{$m^*-1.64\pm0.02/-1.73\pm0.02/-1.90\pm0.01$}  for disturbed/diverse/relaxed samples.  Following Z20, we explore the BCG luminosity evolution with redshift and, thanks to the large mass range in our eRASS1 subsample ($10^{14}\lesssim M_{500} / {\rm M}_\odot\lesssim 10^{15}$), we include a cluster mass dimension. We separate the cluster sample by dynamical state (following \S~\ref{sec:dynamicalstate}) in two mass bins (separated at $M_{500} = 10^{14.4} $M$_\odot$) and two redshift bins (separated at  $z=\textcolor{black}{0.5}$). The cumulative eRASS1 clusters' BCG luminosity distributions,  for the three cluster samples in four bins, are shown in Fig.~\ref{fig:BCGluminosity}. 
For the whole mass and redshift range the BCGs in relaxed clusters are brighter than in disturbed clusters and the diverse population. \textcolor{black}{Similar findings have been reported in the literature for low-redshift clusters \citep[e.g.,][]{wenhan13,lauer14,wenhan15}.} 

\section{Conclusions}
\label{sec:conclusions}
We have used the first eROSITA public data release to classify the dynamical state of \NclusRASS\ clusters and groups,
to study their \BCGXd\ offset distribution and its impact on the BCG luminosity. We used several quality cuts and visual inspection and reduce\textcolor{black}{d} the sample to \NclusRASSusedvisual\ clusters. To our knowledge, this is the largest sample of dynamically classified   \textcolor{black}{X--ray selected}  galaxy clusters available so far.

We model the \BCGXd\ offset distribution using two Rayleigh distributions, representing the relaxed and disturbed sample, as a function of redshift. From the model's parameters (eqs.~\ref{eq:s15} \& \ref{eq:gaussians_expansion}) we find that the $\rho$ remains constant with redshift (meaning that the fraction of relaxed clusters seems to not evolve with time\textcolor{black}{, which is consistent with \cite{ghirardini22} findings}), \textcolor{black}{and mass}.  
We find that  $\sigma_0$ evolves with redshift \textcolor{black}{and mass}, while $\sigma_1$ seems to primarily evolve with mass. In other words, the offset between the BCG and the X--ray peak, in relaxed clusters,  widens with redshift \textcolor{black}{and for lower mass clusters}, showing how central galaxies \textcolor{black}{sink to the bottom of the potential well while the cluster grows in mass over time.}

In the case of disturbed clusters, the BCGs tend to have the largest offset to their X--ray peaks in low-mass clusters, most likely due to the shallow potential well.  
If we used the optical center instead of the BCG, we find similar results; Table~\ref{tab:params} shows the same evolutionary trends for $\sigma_0$ and $\sigma_1$, independently of the center used. Nevertheless, the trends are weaker when the optical center is used. \textcolor{black}{Differences could arise from the central galaxy (CG) selection; redMaPPer CG is determined based on more parameters in addition to brightness and color, including photometric redshift and the local cluster galaxy density around the CG. Selecting the CG using information about the local galaxy density may be problematic for merging clusters of similar richness \citep[see, for example, SPT-CL J0307-6225; ][]{hernandez-lang22}, which could dilute the signal of merging cluster rates.}

To explore the impact of the dynamical state on the BCG, we separate the parent sample in relaxed (\BCGXd<0.25\Rfive), disturbed (\BCGXd>0.5\Rfive), and mixed (diverse)  (0.25<\Rfive\BCGXd<0.5\Rfive) populations. We further separate the sample \textcolor{black}{into} cluster mass and cluster redshift bins. \textcolor{black}{The BCG luminosity distribution can be seen in Fig~\ref{fig:BCGluminosity}.}

We find BCGs in relaxed clusters to be brighter than BCGs in the other two samples at all redshifts. The Kolmog\'orov-Smirnov \textcolor{black}{(KS)} {\it p-value} between the \textcolor{black}{relaxed} and disturbed samples is under \textcolor{black}{0.003}, for all mass and redshift bins. \textcolor{black}{This would imply that losing access to the cluster gas affects BCG growth.  This offset is most evident for the most massive clusters. Furthermore, a comparison of the relaxed bins to each other, produce\textcolor{black}{s} small KS {\it p-values}, while larger values among the disturbed bins.  This indicates that little evolution is observed for BCGs in disturbed clusters as a function of mass and redshift, while the opposite is true for BCGs in the relaxed sample bins. This indicates that the connection to the cluster gas reservoir appears to be key for BCG growth.}

\section*{Acknowledgements}

This project used public archival data from the Dark Energy Survey. 
Funding for the DES Projects has been provided by the US Department of Energy, the US National Science Foundation, the Ministry of Science and Education of Spain, the Science and Technology Facilities Council of the United Kingdom, the Higher Education Funding Council for England, the National Center for Supercomputing Applications at the University of Illinois at Urbana-Champaign, the Kavli Institute of Cosmological Physics at the University of Chicago, the Center for Cosmology and Astro-Particle Physics at the Ohio State University, the Mitchell Institute for Fundamental Physics and Astronomy at Texas A\&M University, Financiadora de Estudos e Projetos, Funda\c c\~ao Carlos Chagas Filho de Amparo \`a Pesquisa do Estado do Rio de Janeiro, Conselho Nacional de Desenvolvimento Cient\'ifico e Tecnol\'ogico and the Minist\'erio da Ci\^encia, Tecnologia e Inova\c c\~ao, the Deutsche Forschungsgemeinschaft and the Collaborating Institutions in the Dark Energy Survey.
The Collaborating Institutions are Argonne National Laboratory, the University of California at Santa Cruz, the University of Cambridge, Centro de Investigaciones Energ\'eticas, Medioambientales y Tecnol\'ogicas-Madrid, the University of Chicago, University College London, the DES-Brazil Consortium, the University of Edinburgh, the Eidgen\"ossische Technische Hochschule (ETH) Z\"urich, Fermi National Accelerator Laboratory, the University of Illinois at Urbana-Champaign, the Institut de Ci\'encies de l'Espai (IEEC/CSIC), the Institut de F\'isica d'Altes Energies, Lawrence Berkeley National Laboratory, the Ludwig-Maximilians Universit\"at M\"unchen and the associated Excellence Cluster Universe, the University of Michigan, the National Optical Astronomy Observatory, the University of Nottingham, The Ohio State University, the University of Pennsylvania, the University of Portsmouth, SLAC National Accelerator Laboratory, Stanford University, the University of Sussex, Texas A\&M University, and the OzDES Member- ship Consortium. Based in part on observations at Cerro Tololo Inter-American Observatory, National Optical Astronomy Observatory, which is operated by the Association of Universities for Research in Astronomy (AURA) under a cooperative agreement with the National Science Foundation.

\textcolor{black}{This work is based on data from eROSITA, the soft X-ray instrument aboard SRG, a joint Russian-German science mission supported by the Russian Space Agency (Roskosmos), in the interests of the Russian Academy of Sciences represented by its Space Research Institute (IKI), and the Deutsches Zentrum f{\"{u}}r Luft und Raumfahrt (DLR). The SRG spacecraft was built by Lavochkin Association (NPOL) and its subcontractors and is operated by NPOL with support from the Max Planck Institute for Extraterrestrial Physics (MPE). 
The development and construction of the eROSITA X-ray instrument were led by MPE, with contributions from the Dr. Karl Remeis Observatory Bamberg \& ECAP (FAU Erlangen-Nuernberg), the University of Hamburg Observatory, the Leibniz Institute for Astrophysics Potsdam (AIP), and the Institute for Astronomy and Astrophysics of the University of T{\"{u}}bingen, with the support of DLR and the Max Planck Society. The Argelander Institute for Astronomy of the University of Bonn and the Ludwig Maximilians Universit{\"{a}}t Munich also participated in the science preparation for eROSITA. 
E. Bulbul, A. Liu, V. Ghirardini, X. Zhang acknowledge financial support from the European Research Council (ERC) Consolidator Grant under the European Union’s Horizon 2020 research and innovation program (grant agreement CoG DarkQuest No 101002585). 
German eROSITA consortium.}

\bibliographystyle{aa} 
\bibliography{new_references} 
\end{document}